\documentclass[prd,aps,amsmath,amssymb,nofootinbib,preprintnumbers]
{revtex4}

\usepackage{graphicx}
\usepackage{dcolumn}
\usepackage{bm}
\usepackage{amsmath}
\usepackage{amsthm}
\usepackage{amsfonts}
\usepackage{euscript,bbm}
\usepackage{ifthen}
\usepackage{psfrag}
\usepackage{slashed}
\usepackage{hyperref}
\usepackage{color}

\newcommand{\degree}{^\circ}

\def\ls{\mathrel{\lower4pt\vbox{\lineskip=0pt\baselineskip=0pt
           \hbox{$<$}\hbox{$\sim$}}}}
\def\gs{\mathrel{\lower4pt\vbox{\lineskip=0pt\baselineskip=0pt
           \hbox{$>$}\hbox{$\sim$}}}}
\def\drawbox#1#2{\hrule height#2pt
\hbox{\vrule width#2pt height#1pt \kern#1pt
              \vrule width#2pt}
              \hrule height#2pt}

\def\Asym#1#2{\vcenter{\vbox{\drawbox{#1}{#2}
              \kern-#2pt       
              \drawbox{#1}{#2}}}}

\newcommand{\be}{\begin{equation}}
\newcommand{\ee}{\end{equation}}
\newcommand{\bea}{\begin{eqnarray}}
\newcommand{\eea}{\end{eqnarray}}

\newcommand{\gsim}{\lower.7ex\hbox{$\;\stackrel{\textstyle>}{\sim}\;$}}
\newcommand{\lsim}{\lower.7ex\hbox{$\;\stackrel{\textstyle<}{\sim}\;$}}

\newcommand{\pythia}{{\tt PYTHIA}}

\newcommand{\ben}{\begin{enumerate}}
\newcommand{\een}{\end{enumerate}}
\newcommand{\bei}{\begin{itemize}}
\newcommand{\eei}{\end{itemize}}

\begin{document}

\title{Dark matter indirect detection signals and the nature of neutrinos in the supersymmetric $U(1)_{B-L}$ extension of the standard model}

\author{Rouzbeh Allahverdi$^{1}$}
\author{Sheldon S. Campbell$^{2,3}$}
\author{Bhaskar Dutta$^{4}$}
\author{Yu Gao$^{4}$}

\affiliation{$^{1}$~Department of Physics and Astronomy, University of New Mexico, Albuquerque, NM 87131 \\
$^{2}$~Center for Cosmology and AstroParticle Physics, Ohio State University, Columbus, OH, 43210 \\
$^{3}$~Department of Physics, Ohio State University, Columbus, OH 43210 \\
$^{4}$~Department of Physics and Astronomy, Mitchell Institute for Fundamental Physics and Astronomy, Texas A\&M University, College Station, TX 77843-4242
}

\begin{abstract}
In this paper, we study the prospects for determining the nature of neutrinos in the context of a supersymmetric $B-L$ extension of the standard model by using dark matter indirect detection signals and bounds on $N_{\text{eff}}$ from the cosmic microwave background data. The model contains two new dark matter candidates whose dominant annihilation channels produce more neutrinos than neutralino dark matter in the minimal supersymmetric standard model.
The photon and neutrino counts may then be used to discriminate between the two models. If the dark matter comes from the B-L sector, its indirect signals and impact on the cosmic microwave background can shed light on the nature of the neutrinos. When the light neutrinos are of Majorana type, the indirect neutrino signal from the Sun and the galactic center may show a prompt neutrino box-feature, as well as an earlier cut-off in both neutrino and gamma ray energy spectra.
When the light neutrinos are of Dirac type, their contribution to the effective number of neutrinos $N_{\text{eff}}$ is at a detectable level.
\end{abstract}
MIFPA-14-18

\maketitle

\section{Introduction}

There are various lines of evidence supporting the existence of dark matter (DM) in the Universe, but its identity remains a major problem at the interface of particle physics and cosmology. The weakly interacting massive particles (WIMPs), which typically arise in extensions of the standard model (SM), are promising candidates for DM in the Universe~\cite{bhs}. There are currently major direct, indirect, and collider experimental searches underway to detect WIMP-like DM and determine its properties.

Another great mystery in particle physics is the origin of neutrino mass. Explaining neutrino mass and mixing, which are observed in solar and atmospheric neutrino oscillations, has been the focus of extensive theoretical investigations in beyond the SM physics~\cite{bib:bsm}. The nature of neutrinos (whether they are their own antiparticles or not) is also an important question whose answer will likely have major consequences for generation of matter-antimatter asymmetry in the universe~\cite{lepto}.

In an interesting class of models, the DM candidate originates from the same sector that is responsible for neutrino mass and mixing (for some explicit examples, see~\cite{link}). Such a link provides potential opportunities in the quest for understanding these two great mysteries in fundamental physics. In particular, DM detection experiments--in addition to providing information about DM and its parameters--may in principle be used to shed light on neutrinos and their properties.

In this paper, we consider a minimal and well-motivated supersymmetric extension of the SM as a case study to demonstrate this possibility. The model includes a gauged $U(1)_{B-L}$ symmetry~\cite{mm} (where $B$ and $L$ are baryon and lepton number respectively). Anomaly cancellation then implies the existence of three right-handed (RH) neutrinos and allows us to write the Dirac and Majorana mass terms for the neutrinos to explain the mass and mixing of light neutrinos. This model provides two new DM candidates both arising from the $B-L$ extension: the lightest neutralino in the $B-L$ sector and the lightest RH sneutrino.

The goal of our study is two-fold. First, we investigate the possibility of using indirect detection signals to distinguish the new DM candidates from the extensively studied lightest supersymmetric particle (LSP) in the minimal supersymmetric standard model (MSSM). Second, we explore the possibility to identify the nature of neutrinos (Majorana vs. Dirac) in the $B-L$-extended MSSM once this model is differentiated from the MSSM.

We achieve the first goal by using the dominant DM annihilation final states in the $B-L$ model as a discriminator. The presence of neutrino-philic or prompt neutrino final states in the $B-L$ model result in indirect signals that are distinct from those in the MSSM. The ratio of the neutrino to gamma-ray fluxes from DM annihilation in the galactic center is one way to establish a neutrino-philic final state. Moreover, the shape of the spectrum of the neutrino signal from DM annihilation in the galactic center and the Sun may contain a novel prompt-neutrino final state, which can only be present if neutrinos are Majorana. To achieve the second goal, we use the effective number of neutrinos $N_{\rm eff}$ as a complementary probe. $N_{\rm eff}$, which is tightly constrained by the cosmic microwave background (CMB) experiments, can discriminate between Majorana and Dirac neutrinos in the $B-L$ model in a robust manner regardless of the dominant DM annihilation final state.

The paper is organized as follows. In Section II we describe the $U(1)_{B-L}$ extension of MSSM, the new DM candidates that it contains, and prospects for their direct detection. In Section III we discuss the gamma-ray and neutrino signals from annihilation of the $B-L$ DM candidates from the galactic center, and the neutrino signal from gravitationally trapped DM particles inside the Sun in detail. In Section IV we show how $N_{\rm eff}$ can be used to convincingly distinguish between Majorana and Dirac neutrinos in the $B-L$ model. We close in Section V by presenting a summary discussion and our conclusion.

\section{Dark Matter in $U(1)_{B-L}$ Extension of MSSM}

\subsection{The Model}

We consider an extension of the MSSM that includes a gauged $U(1)_{B-L}$ and conserves $R$-parity. The $B-L$ extension of the SM gauge group is very well motivated~\cite{mm} since, as mentioned above, it automatically implies the existence of three RH neutrinos through which one can explain the neutrino mass and mixing.

The model contains a new gauge boson $Z^{\prime}$, two new Higgs fields $H^{\prime}_1$ and $H^{\prime}_2$, three RH neutrinos $N$, and their supersymmetric partners. The superpotential is given by (the boldface characters denote superfields):
\begin{equation} \label{sup}
W = W_{\rm MSSM} + W_{B-L} + y_D {\bf N} {\bf H_u} {\bf L}  \, ,
\end{equation}
where ${\bf H_u}$ and ${\bf L}$ are the superfields containing the Higgs field that gives mass to up-type quarks and the left-handed (LH) leptons respectively. For simplicity, we have omitted the family indices. The last term on the RH side of Eq.~(\ref{sup}) is the neutrino Yukawa coupling term. The $B-L$ charges of leptons, quarks, $N$, $H^{\prime}_1$, and $H^{\prime}_2$ are in general given by $Q_L$, $-Q_L/3$, $Q_L$, $Q$, and $-Q$ respectively. The MSSM Higgs fields have zero $B-L$ charge.

The $U(1)_{B-L}$ symmetry is broken by the vacuum expectation value (VEV) of $H^{\prime}_1$ and $H^{\prime}_2$, denoted by $v^{\prime}_1$ and $v^{\prime}_2$ respectively, with ${\rm tan} \beta^{\prime} \equiv v^{\prime}_2/v^{\prime}_1$. This results in a mass $m_{Z^{\prime}} = g_{B-L} Q \sqrt{{v^{\prime}_1}^2 + {v^{\prime}_2}^2}$ for the $Z^{\prime}$ gauge boson. Three physical Higgs fields $\phi, ~ \Phi$ (scalars) and ${\cal A}$ (a pseudo scalar) are left in the $B-L$ sector spontaneous breakdown of $U(1)_{B-L}$.

The $W_{B-L}$ term in Eq.~(\ref{sup}) contains ${\bf H^{\prime}_1}$, ${\bf H^{\prime}_2}$, and ${\bf N^c}$. Its detailed form depends on the charge assignments of the $B-L$ Higgs fields. For example, if $Q = 2 Q_L$, we can have:
\begin{equation} \label{b-lsup}
W_{B-L} = \mu^{\prime} {\bf H^{\prime}_1} {\bf H^{\prime}_2} + {1 \over 2} y_M {\bf H^{\prime}_1} {\bf N^c} {\bf N^c} .
\end{equation}
The first term on the RH side of this equation is the $B-L$ Higgs $\mu$-term, and the second term is a neutrino Majorana Yukawa coupling that generates a Majorana mass $M_N = y_M v^{\prime}_1$ for the RH neutrinos after $U(1)_{B-L}$ breaking. If $Q \neq \pm 2 Q_L$, no Majorana Yukawa coupling (neither to $H^{\prime}_1$ nor to $H^{\prime}_2$) will be allowed. In this case, neutrinos will be Dirac fermions.

\subsection{Dark Matter Candidates}

The $U(1)_{B-L}$ model has two new dark matter (DM) candidates. The first candidate, denoted by ${\tilde \chi}^0_1$, is the lightest neutralino in the $B-L$ sector~\cite{khalil,pamela1}. It is a linear combination of the $U(1)_{B-L}$ gaugino ${\tilde Z}^{\prime}$ and the two Higgsinos ${\tilde H}^{\prime}_1$ and ${\tilde H}^{\prime}_2$. The other candidate, denoted by ${\tilde N}$, is the lightest of the RH sneutrinos~\cite{bhaskar,pamela2}.\footnote{It is also possible to have viable sneutrino DM without the inclusion of $U(1)_{B-L}$ symmetry. For example, see~\cite{munoz} for RH sneutrino DM and~\cite{fornengo,march-russell,belanger} for mixed sneutrino DM.} If supersymmetric particles in the MSSM sector are heavier than those in the $U(1)_{B-L}$ sector, the DM will be either ${\tilde X}^0_1$ or ${\tilde N}$, whose stability is ensured by $R$-parity. These DM candidates can obtain the correct relic abundance via thermal freeze-out if the scale of $U(1)_{B-L}$ breaking is not much above TeV.

\begin{figure}
\includegraphics[scale=0.6]{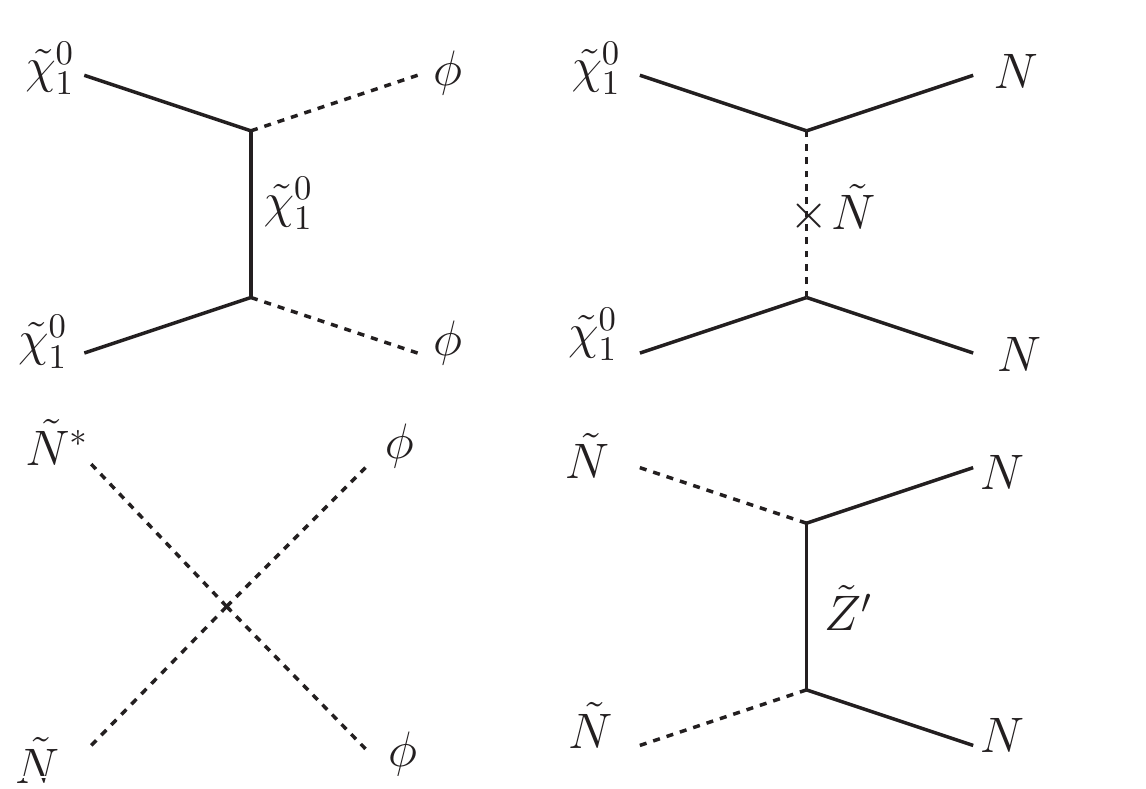}
\caption{Leading Feynman diagrams for $S$-channel annihilation of DM in the $U(1)_{B-L}$ model: $B-L$ neutralino DM (top) and RH sneutrino DM (bottom).}
\label{fig:feyn}
\end{figure}

Each of the $B-L$ DM candidates has various DM annihilation channels. However, since we are interested in indirect signals from DM annihilation, we focus on those channels with $S$-wave dominance as they make the largest contributions to annihilation at the present time. The most relevant final states in this category are the lightest $B-L$ Higgs $\phi$~\cite{pamela1,pamela2} and the RH neutrino $N$~\cite{icecube1}. The leading annihilation diagrams for these final states are shown in Fig.~(\ref{fig:feyn}).

\vskip 3mm
\noindent
{\bf (1) ${\rm DM} + {\rm DM} \rightarrow N N$:}
\\
\noindent
$\bullet$ ${\tilde \chi}^0_1 {\tilde \chi}^0_1 \rightarrow N N$: If ${\tilde \chi}^0_1$ is the DM candidate, the annihilation proceeds via the $t$-channel exchange of ${\tilde N}$ with mass flipping. We see from Eq.~(\ref{b-lsup}) that ${\tilde N}$ mass flipping arises from the Majorana mass term and soft supersymmetry breaking. Therefore the $N N$ final state is only possible if neutrinos are Majorana fermions.
\\
\noindent
$\bullet$ ${\tilde N} {\tilde N} \rightarrow N N$: If ${\tilde N}$ is the DM candidate, the annihilation proceeds via the $t$-channel exchange of ${\tilde Z}^{\prime}$ with soft supersymmetry breaking mass of ${\tilde Z}^{\prime}$ inserted. In this case the $N N$ final state arises for both Majorana and Dirac neutrinos.
\vskip 2mm
\noindent
If neutrinos are Majorana fermions, each of the RH neutrinos $N$ in turn decays to the lightest MSSM Higgs $h$ and a LH neutrino $\nu$.~\footnote{Assuming that the RH neutrino Majorana mass is $\sim 1$ TeV, which happens if $U(1)_{B-L}$ is broken around TeV, $y_D \lesssim 10^{-6}$ is required in order to obtain light neutrino mass of $\sim {\cal O}(0.1 ~ {\rm eV})$. This results in a very short life time $\sim 10^{-13}$ s for $N$ decay, which implies the decay is instant and occurs at the same place as DM annihilation.} Subsequent decay of the Higgs $h$ primarily produces a $b {\bar b}$ pair. If neutrinos are Dirac fermions, the $N \rightarrow H \nu$ decay is blocked since $N$ has the same mass as $\nu$. $N$ interaction with other particles is via $Z^{\prime}$ exchange, which is much more suppressed than the weak interactions due to the bound $(m_Z^{\prime}/g_{B-L} Q_L) > 6$ TeV from Tevatron and LEP~\cite{Z'}. This implies that the $N N$ final state will go undetected in the case of Dirac neutrinos.
\vskip 2mm
\noindent
We conclude that DM + DM $\rightarrow N N$ annihilation channel mainly produces the $2 \nu + 4b$ final state in case of Majorana neutrinos. The $b$ quarks make the main contribution to the gamma-ray signal, while the neutrino signal is mainly due to $\nu$'s in the final state. If neutrinos are Dirac, the $N N$ annihilation channel will not be visible.

{\it {The signal arising from DM annihilation depends on the nature of neutrinos in this case. Therefore, indirect detection may be used to identify the nature of neutrinos.}}
\vskip 3mm
\noindent
{\bf (2) ${\rm DM} + {\rm DM} \rightarrow \phi \phi$:}
\\
\noindent
$\bullet$ ${\tilde \chi}^0_1 {\tilde \chi}^0_1 \rightarrow \phi \phi$: If ${\tilde \chi}^0_1$ is the DM candidate, the annihilation proceeds via the $t$-channel exchange of ${\tilde \chi}^0_1$.
\\
\noindent
$\bullet$ ${\tilde N} {\tilde N}^* \rightarrow \phi \phi$: If ${\tilde N}$ is the DM candidate, the annihilation proceeds via interactions in the $D$-term of the scalar potential. The relevant interaction term is~\cite{pamela2}:
\begin{equation} \label{dterm}
V \supset {1 \over 2} g^2_{B-L} Q Q_L ~ {\rm cos} (2 \alpha^{\prime}) |{\tilde N}|^2 \phi^2 ,
\end{equation}
where $\alpha^{\prime}$ is the mixing angle that relates the $B-L$ scalar Higgses $\phi,~\Phi$ to $H^{\prime}_1,~H^{\prime}_2$.
\vskip2mm
\noindent
Each of the $\phi$ particles decays to leptons and quarks and the corresponding branching fractions are~\cite{pamela1,pamela2}:
\begin{eqnarray} \label{phidecay}
&& {\rm Br}_{\phi \rightarrow l {\bar l}} \propto Q^4_L \left({m_l \over m_{\rm DM}}\right)^2 \, , \nonumber \\
&& {\rm Br}_{\phi \rightarrow q {\bar q}} \propto 3 \cdot 3^{-4} \cdot Q^4_L \left({m_q \over m_{\rm DM}}\right)^2 \, .
\end{eqnarray}

In the second expression, the factor of $3$ is due to color multiplicity for quarks, the factor of $3^{-4}$ is due to $B-L$ charge assignments of leptons and quarks. The factors $(m_l/ m_{\rm DM})^2$ and $(m_q/m_{\rm DM})^2$ are due to mass suppression of the decay ($m_l$ and $m_q$ being the mass of the produced lepton and quark respectively). This implies that $\phi$ preferentially decays to the heaviest lepton and quark. If $m_\phi < 2 m_t$ (with $m_t$ being the top quark mass), the kinematically allowed decay modes are $\phi \rightarrow \tau {\bar \tau}$ and $\phi \rightarrow b {\bar b}$, where:
\begin{equation} \label{br}
{\rm Br}_{\phi \rightarrow \tau {\bar \tau}} \approx 7 {\rm Br}_{\phi \rightarrow b {\bar b}}.
\end{equation}
\vskip 2mm
\noindent
We conclude that DM + DM $\rightarrow \phi \phi$ annihilation channel mainly produces the $4 \tau$ final states regardless of the Majorana or Dirac nature of neutrinos. Both of the gamma-ray and neutrino signals arise from $\tau$'s in the final state.

{\it{The signal arising from DM annihilation does not depend on the nature of neutrinos in this case.}}

\subsection{Direct Detection Prospects}

The interactions between either of the DM candidates, ${\tilde \chi}^0_1$ or ${\tilde N}$, and quarks arise from the $B-L$ sector. Therefore, since the $U(1)_{B-L}$ symmetry is vectorial (i.e., the RH and LH quarks have the same $B-L$ charge), we strictly have spin-independent (SI) interactions between DM and nucleons in this model. In below, we discuss the cases with ${\tilde \chi}^0_1$ and ${\tilde N}$ DM separately.
\vskip 2mm
\noindent
{\bf (1) ${\tilde \chi}^0_1$ as DM:} The leading order interaction between ${\tilde \chi}^0_1$ and quarks is via squark exchange in the $t$-channel. There are also interactions via $\phi,~\Phi,~{\cal A}$ exchange, but these are much suppressed because the $B-L$ Higgs fields couple to the quarks at one-loop level.

For ${\cal O}({\rm TeV})$ squark masses, the DM-nucleon scattering cross section is found to be $\sigma_{\rm SI} \ll 10^{-10}$ pb~\cite{pamela1}, which is well beyond the reach of direct detection experiments in the near future.
\vskip 3mm
\noindent
{\bf (2) ${\tilde N}$ as DM:} The leading interaction between ${\tilde N}$ and quarks is via $Z^{\prime}$ exchange in the $t$-channel. This results in the following SI scattering cross section~\cite{icecube1}:
\begin{equation} \label{sigmaN}
\sigma_{\rm SI} \approx {1 \over 8 \pi} \left({g_{B-L} Q_L \over m_{Z^{\prime}}}\right)^4 m^2_n ,
\end{equation}
where $m_n$ is the nucleon mass.

The tightest experimental bounds on $\sigma_{\rm SI}$ are currently provided by the LUX experiment~\cite{lux}. The most stringent limit $\sigma_{\rm SI} \lesssim 10^{-45}$ cm$^2$ arises for DM mass around 40 GeV, which is translated to $(m_{Z^\prime}/g_{B-L} Q_L) \gtrsim 10$ TeV. This is stronger than the collider bound of $(m_{Z^{\prime}}/g_{B-L} Q_L) > 6$ TeV from Tevatron and LEP~\cite{Z'}. We note, however, that the LUX limit becomes weaker by a factor of few for DM masses above 100 GeV, which makes it comparable to the existing collider bound.

An important point to note is that the ${\tilde N}-{\tilde N}-Z^{\prime}$ interaction vertex involves real and imaginary parts of ${\tilde N}$, denoted by $N_{\rm R}$ and ${\tilde N}_{\rm I}$ respectively. This implies that DM-nucleon scattering can occur only if ${\tilde N}_{\rm R}$ and ${\tilde N}_{\rm I}$ are almost degenerate in mass. More precisely, the mass difference needs to be smaller than the DM kinetic energy, so that ${\tilde N}_{\rm R}$ scattering to ${\tilde N}_{\rm I}$ (and vice versa) can occur in the detector.\footnote{${\tilde N}_{\rm R} \rightarrow {\tilde N}_{\rm R}$ scattering occurs at one-loop level, and hence is highly suppressed.} The situation then depends on whether $N$ has a Dirac or Majorana nature.
\vskip 2mm
\noindent
$\bullet$ If ${\tilde N}$ is the superpartner of the RH component of a Dirac neutrino, its mass is given by:
\begin{equation} \label{snmassd}
m^2_{\tilde N} = m^2_0 + {1 \over 4} m^2_{Z^{\prime}} {\rm cos} 2 \beta^{\prime} ,
\end{equation}
where $m_0$ is the soft supersymmetry breaking mass of ${\tilde N}$. In this case ${\tilde N}_{\rm R}$ and ${\tilde N}_{\rm I}$ have exactly the same mass, which implies that the DM candidate is a complex scalar. As a result, ${\tilde N}$-nucleon scattering via $Z^{\prime}$ exchange proceeds and the above discussion is directly applied in this case.
\vskip 2mm
\noindent
$\bullet$ If ${\tilde N}$ is the superpartner of a Majorana neutrino, see Eq.~(\ref{b-lsup}), its mass is given by:
\begin{equation} \label{snmassm}
m^2_{\tilde N_{\rm R,I}} = m^2_0 + m^2_N \pm (B_N + \mu^{\prime} {\rm tan} \beta^{\prime}) m_N + {1 \over 4} m^2_{Z^{\prime}} {\rm cos} 2 \beta^{\prime} ,
\end{equation}
where $m_N$ is the Majorana mass and $B_N$ is its associated $B$-term. It is seen that the third term on the RH side of the above expression splits the masses of the real and imaginary parts of $\tilde N$. The DM candidate is the lighter field, taken to be ${\tilde N}_{\rm R}$, which is a real scalar. The $t$-channel exchange of $Z^{\prime}$ between the DM and nucleons then results in upscattering of ${\tilde N}_{\rm R}$ to ${\tilde N}_{\rm I}$, which can happen only if $|m_{{\tilde N}_{\rm R}} - m_{{\tilde N}_{\rm I}}| < {\cal O}({\rm MeV})$. Whether or not this is the case is a model-dependent issue.
%
\vskip 3mm
\noindent
We conclude that prospects for direct detection of DM candidates in the $B-L$ model depends on their identity as well as the nature (Majorana or Dirac) of neutrinos. If ${\tilde N}$ is the DM particle, and its fermionic partner $N$ has Dirac nature, then $\sigma_{\rm SI}$ can be just below the current experimental limits. On the other hand, if $N$ has Majorana nature, then its detection prospects depend on model details of may be just around the corner or it may completely escape detection. If ${\tilde \chi}^0_1$ is the DM particle, then $\sigma_{\rm SI}$ is considerably below the current experimental limits, and hence no realistic prospect for its direct detection.

\section{Indirect Detection} \label{sect:ID}

We have established that, in the context of the gauged $U(1)_{B-L}$ extension of the MSSM, when the DM particle is the lightest neutralino from the $B-L$ sector ${\tilde \chi}^0_1$ or the lightest RH sneutrino ${\tilde N}$, then some important annihilation channels are not observed if the neutrinos are Dirac fermions. In this section, we explore MSSM and $B-L$-extension dark matter models with observable indirect detection signals, and present the different signals. We compare the different indirect detection signals in more detail in Section~\ref{sect:discussion}.

\subsection{Signals from Galactic Dark Matter Annihilation}
The annihilation of the DM candidates inside galactic halos can yield gamma-ray and neutrino signals that can be probed by the Fermi-LAT and IceCube experiments. The prompt gamma-ray and neutrino spectrum from an annihilation event is convoluted with the line-of-sight cosmic dark matter distribution, specified for cosmologically local annihilations (assumed to be dominantly through an S-wave interaction) by the J-factor $J(\vec{\theta}\,)$ along the direction $\vec{\theta}$ of interest,
\begin{equation}
J(\vec{\theta}\,)=\int ds\,\rho^2(s,\vec{\theta}\,),
\end{equation}
with $\rho$ the DM density at distance $s$. The differential flux of annihilation products $X$ incident on our position from the solid angle $\Omega$ is
\begin{equation}
\frac{d\phi_X}{dE}=\frac{\left< \sigma v\right>}{8\pi\xi m_{\text{DM}}^2}\frac{dN_X}{dE}\frac{1}{\Omega}\int_\Omega d^2\theta\,J(\vec{\theta}\,),
\end{equation}
where $m_{\text{DM}}$ is the DM particle mass, $\sigma v$ is the relative-velocity weighted s-wave annihilation cross section, $dN_X/dE$ is the prompt annihilation spectrum to product $X$, and $\xi$ is a multiplicity factor. When DM is its own antiparticle, such as for $\tilde{\chi}_1^0$ DM, then $\xi=1$. For $\tilde{N}$ DM, we will assume the scenario where the cosmological abundances of each species $\tilde{N}$ and $\tilde{N}^*$ are equal. In this case, $\xi=2$ for $\tilde{N}\tilde{N}^*$ annihilations, and $\xi=4$ for $\tilde{N}\tilde{N}$ or $\tilde{N}^*\tilde{N}^*$ separately (we can take $\xi=2$ for them combined when they produce equal abundances of $X$). The sample calculations shown in this paper are for $\xi=1$ with a fiducial annihilation cross section of $\left<\sigma v\right> =1\text{ pb}=3\times10^{-26}\text{ cm$^3$/s}$.

In our study, we considered DM masses of 200 GeV, 300 GeV, 400 GeV, and 600 GeV. For each mass, we contrast four different annihilation spectra, from annihilations to $\phi\phi$, $NN$, $W^+W^-$, and $t\bar{t}$. The first two spectra are representative of annihilations of B-L sector dark matter. The $\phi\phi$ channel is to 2 lightest B-L sector Higgs particles, each taken to have mass $m_\phi=100\text{ GeV}$, though the results are insensitive to this value. Each $\phi$ is taken to decay promptly to $\tau^+\tau^-$ with branching ratio $7/8$, and the remaining decays are to $b\overline{b}$. The $NN$ channel is to 2 RH neutrinos (democratically in flavor), each of which decays to a LH neutrino and an $m_h=126\text{ GeV}$ lightest MSSM Higgs boson. In this case, the kinematics of the products, particularly the prompt LH neutrinos, is dependent on the mass of the RH neutrino. For our sample calculations, we chose to consider cases where the mass $m_N$ of the RH neutrino is 15 GeV less than the DM mass.

We contrast these two spectra with more typical annihilations found in the MSSM, to investigate if the $B-L$ DM is distinguishable from a typical MSSM DM. In MSSM, a bino DM also produces $b\overline{b}$ and $\tau^+\tau^-$, but suffers from its cross-section being p-wave suppressed. If the DM has significant wino or higgsino content, then it annihilates predominantly to $W^+W^-$ bosons. When top quarks are a kinematically viable channel, they can also have a large branching ratio, and we consider them in the $t\bar{t}$ spectrum.

Each spectrum was calculated via Monte-Carlo simulation using \pythia~8.180 \cite{pythia}. This includes the final state radiation (FSR) processes of gamma-ray bremsstrahlung from external charged fermion legs, but not from $W$ bosons. This FSR produces a ``box''-shaped spectrum with low flux, and causes a harder gamma-ray spectrum at energies near the DM mass \cite{FSR}. This is an important effect for $m_{\text{DM}}\agt1\text{ TeV}$ annihilations to $W^+W^-$ where showering softens the spectrum considerably, but for our considerations, the FSR effect is not noticeable, and we do not need to account for it. In addition to prompt annihilation radiation, gamma rays and neutrinos are also produced when cosmic-ray annihilation products (such as electrons) interact with the intergalactic medium. These contributions to the observed spectra are sensitive to models of the distribution of galactic baryons, magnetic fields, and the physics of cosmic-ray propagation. These details are beyond the scope of this present work and will not be considered here. We note, however, that the cosmic-ray contributions to the gamma-rays and neutrinos will brighten the spectra at energies significantly lower than the DM mass, and have little effect near the DM mass.

If the DM density profile near the galactic center (GC) of the Milky Way galaxy is well-described by an Einasto cusp \cite{Aquarius}, then it is a leading cosmic source of annihilation radiation, and will be the focus of discussion in this work. The Einasto density profile is
\begin{equation}
\rho(r)=\rho_{-2}\exp\left(-\frac{2}{\alpha_E}\left[\left(\frac{r}{r_{-2}}\right)^{\alpha_E}-1\right]\right).
\end{equation}
We adopt the best-fit Galactic model of the analysis of dynamical tracer data in \cite{Catena} with scale density $\rho_{-2}=0.08\text{ GeV/cm$^3$}$, scale radius $r_{-2}=20\text{ kpc}$, and shape parameter $\alpha_E=0.22$, with the solar radius at $R_\odot=8.25\text{ kpc}$. This halo has a Galactic halo mass of $M=1.47\times10^{12}\ M_\odot$.

\begin{figure}[h]
\includegraphics[scale=0.6]{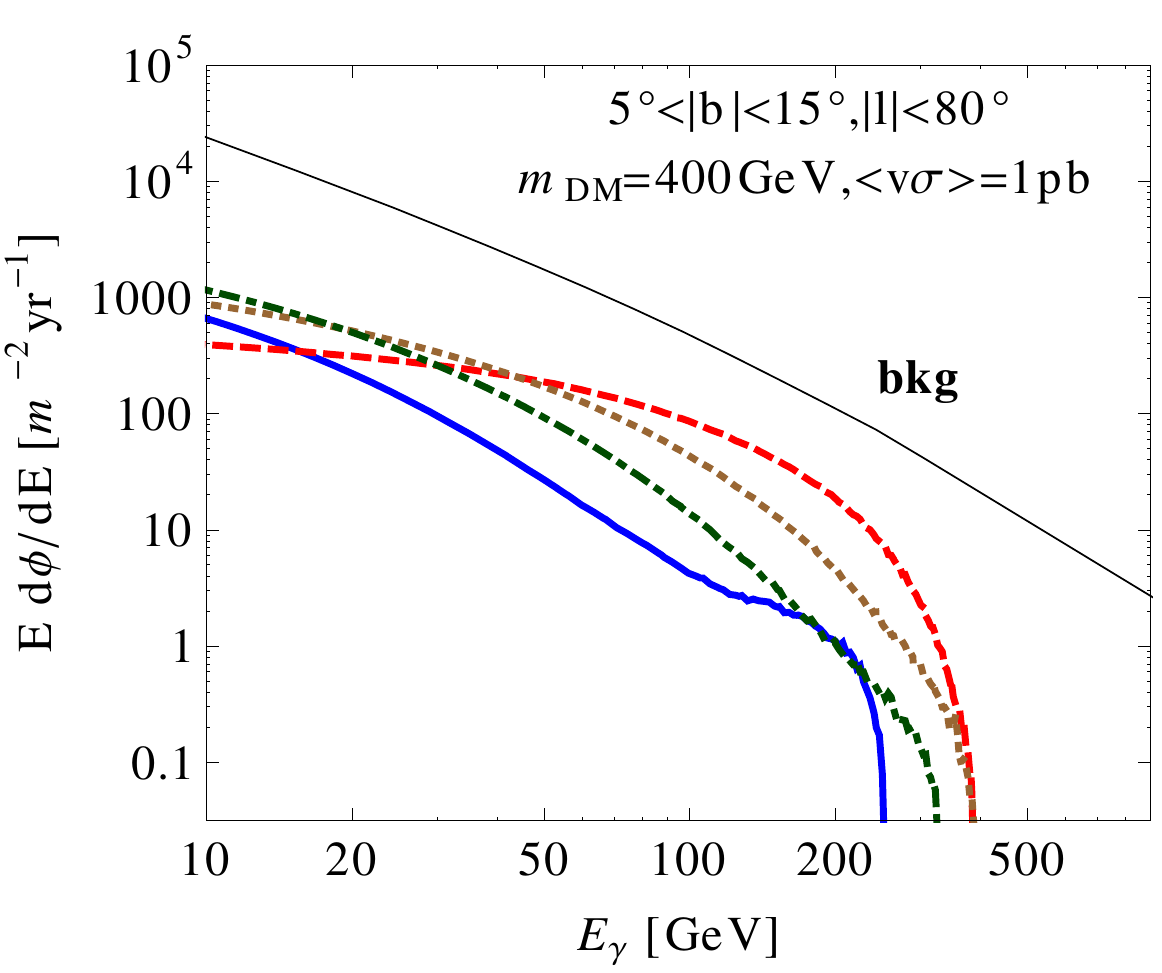}
\caption{The gamma-ray signal spectra from DM annihilation at the galactic center. The red dashed line is the $\phi\phi$ spectrum, yellow dotted is $W^+W^-$, green dot-dashed is $t\bar{t}$, and blue solid is $NN$. We point out a clear hierarchy between the four spectra when E is above 50 GeV, with the $B-L$ sector spectra at the extremes. The $NN$ spectrum is only allowed in $B-L$ DM models if the neutrino is Majorana. The Fermi-LAT background is from Ref.~\cite{Ackermann:2012rg}}
\label{fig:photon_spectra}
\end{figure}

\begin{figure}[h]
\includegraphics[scale=0.5]{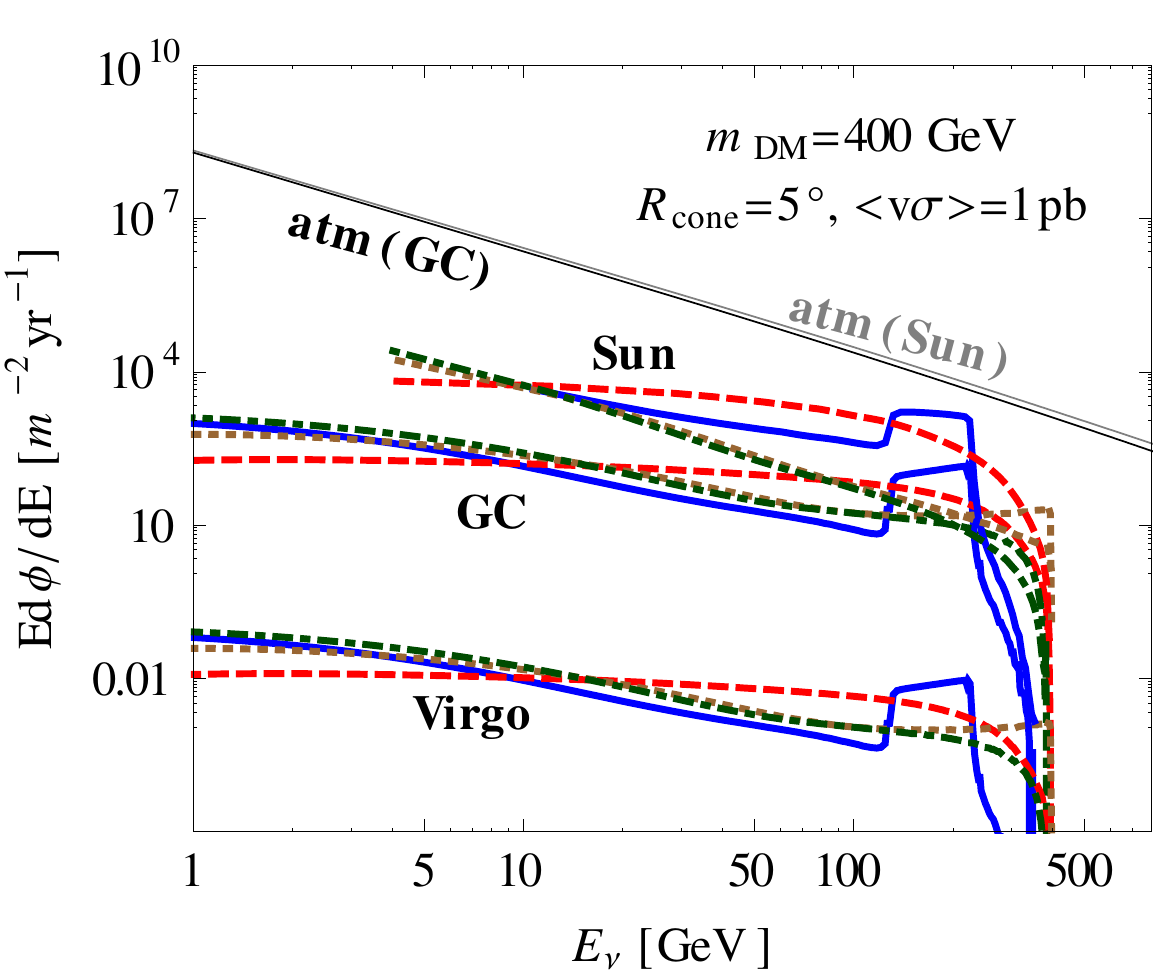}
\includegraphics[scale=0.5]{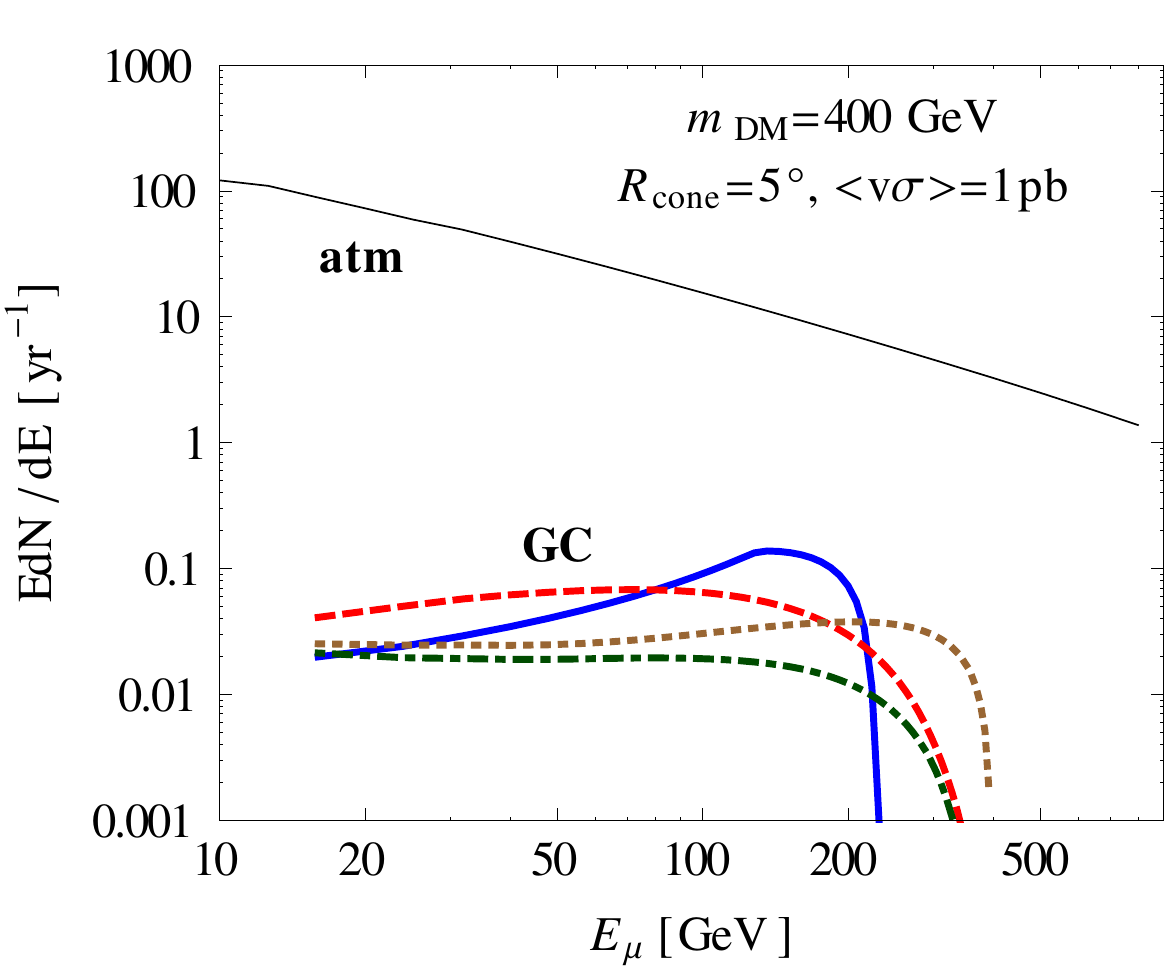}
\includegraphics[scale=0.5]{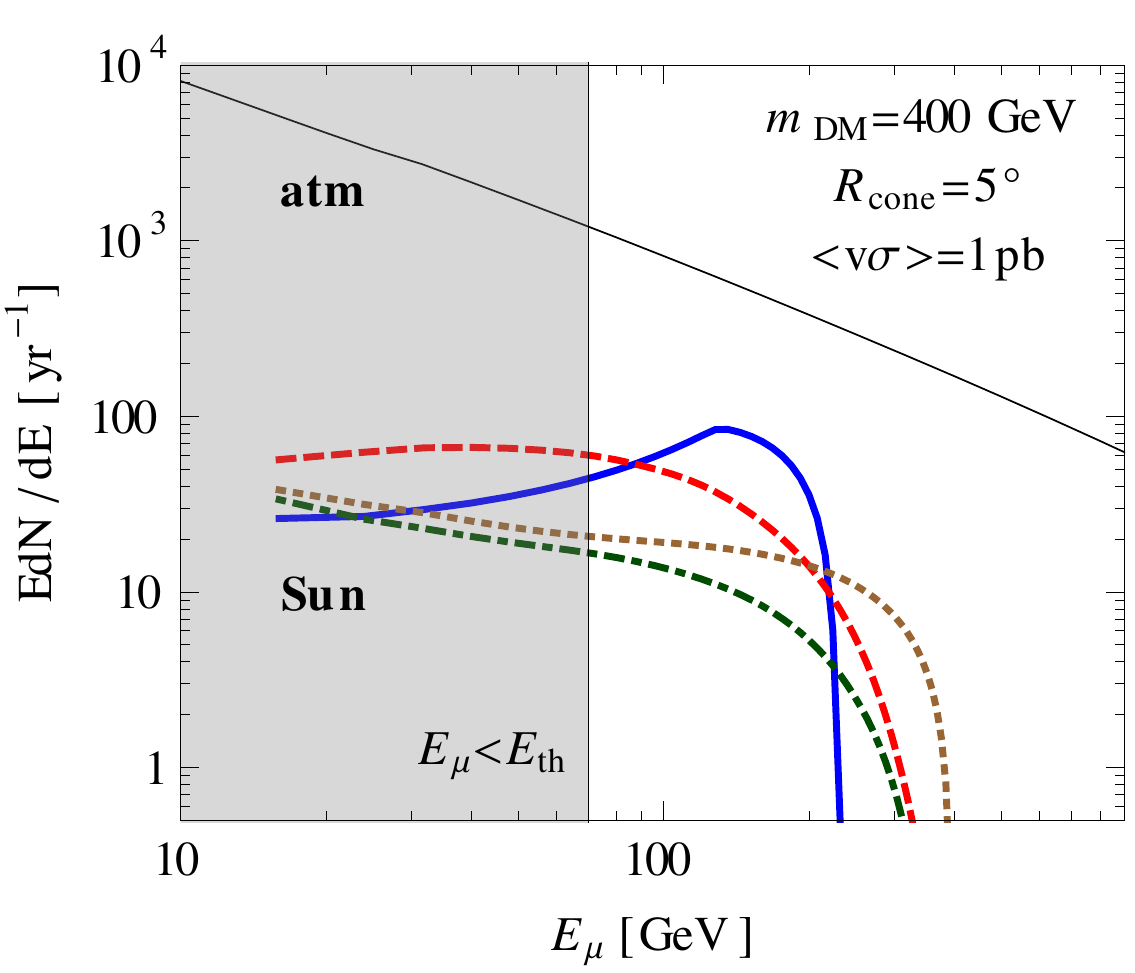}
\caption{Propagated neutrino spectra from the leading galactic sources (left), and the contained $\mu$ event energy spectra (GC only) at the DeepCore (DC) detector (right). The curves are as in Fig.~\ref{fig:photon_spectra}. The atmospheric background curves are in the GC direction. The $NN$ spectrum is only allowed in $B-L$ DM models if the neutrino is Majorana.}
\label{fig:nu_spectra}
\end{figure}

Fig.~\ref{fig:photon_spectra} and ~\ref{fig:nu_spectra} (left panel) illustrate the galactic gamma-ray and neutrino signals, for a 400 GeV DM. For the gamma rays, we used the middle latitude `region of interest' (ROI) window~\cite{Ackermann:2012rg} $5\degree <|b|<15\degree, |l|<80\degree$ to avoid the large astrophysical uncertainty in the galactic central region. For the neutrinos, we focused on the GC and chose a $5^\circ$ radius observation cone to reflect the expected angular resolution of IceCube.

The gamma-ray signals have a clear hierarchy in flux between the four example spectra for $E\agt m_{\text{DM}}/8$, with $\phi\phi$ brightest, followed by $W^+W^-$, $t\bar{t}$, and $NN$. At lower energies, the $\phi\phi$ spectrum begins to fall below the other spectra. This hierarchy is robust for the energy range we considered, except that the $t\bar{t}$ spectrum becomes the weakest when $m_{\text{DM}}\alt 300\text{ GeV}$. A $B-L$-sector DM will then be expected to produce the $\phi\phi$ spectrum, and will be brighter than corresponding MSSM models. The $NN$ spectrum will be disallowed if the $\nu$'s are Dirac, but that would be difficult to ascertain from gamma-ray observations alone.

For neutrinos, we use the latest oscillation parameters~\cite{bib:nu_par} to compute the $\nu_\mu$ component in the signal spectrum that is considered to be fully de-cohered over astronomical distances. Because the GC is above the ground at the South pole, only the DeepCore~\cite{DeYoung:2011ke} effective volume is included (though a larger effective volume is used by IceCube in their analyses), and the derived muon event spectra are shown in the right panel of Fig.~\ref{fig:nu_spectra}. For the atmospheric background, we use the data from the Super-Kamiokande measurement~\cite{Honda:2006qj}.

At the peak number flux of the neutrino spectra, the $t\bar{t}$ signal remains the weakest, but the other three have reversed their hierarchy, with $NN$ coming closest to the atmospheric background. The next closest spectrum to the atmospheric background is $W^+W^-$, and then $\phi\phi$. However, these peaks occur at different energies owing to the different shapes of the spectra. As is well known, the neutrino spectrum from production of $W$ bosons is very hard. Therefore, the soft $\phi\phi$ spectrum is easily distinguishable from both the much harder $W^+W^-$ and much softer $t\bar{t}$. The $NN$ spectrum is very interesting. It contains two components: the very soft spectrum from the decay of the standard model Higgs, and a box-shaped spectrum of the prompt neutrinos. \emph{The presence of the prompt neutrino box feature is a clear feature that can only be present if neutrinos are Majorana.} The energy range for the prompt neutrinos from $N$ decays is
\begin{equation}
E_{\text{prompt}}\in E_\nu\left(\frac{m_{\text{DM}}}{m_N}\pm\sqrt{\left(\frac{m_{\text{DM}}}{m_N}\right)^2-1}\right) ,
\end{equation}
where $E_\nu=(m_N^2-m_h^2)/(2m_N)$ is the energy of the prompt neutrino in the rest frame of the decaying $N$. In Fig.~\ref{fig:nu_spectra}, the prompt neutrino bump from $N$ decay is over 130--227 GeV.

It can be clearly seen that the shape of the gamma ray and neutrino(muon) spectrum differ among the possible annihilation channels, which can potentially be exploited to shed light on the nature of the dark matter candidate, hence whether the neutrino is Majorana or Dirac in this $B-L$ model. In Section~\ref{sect:discussion} we will comment further on these features in the gamma ray and neutrino spectra, give their annual signal rates and apply general cuts to distinguish the different channels.

While the galactic center is a clean environment for the neutrino signal, it contains complicated astrophysical gamma-ray foregrounds. Additionally, the signal will be difficult to observe from the GC if the density profile is not cuspy. Dwarf spheroidal galaxies are a clean environment to see annihilation to gamma-rays, but these sources are too small to be seen with a neutrino telescope's angular resolution. Nearby galaxy clusters are extended sources, making them ideal targets for IceCube, but they also contain complicated astrophysics gamma-rays that may hide DM annihilation.

For reference, we plot in Fig.~\ref{fig:nu_spectra}, left-panel, the signal neutrino flux from the Virgo cluster~\cite{nu_cluster}, whose contribution is subdominant compared to the GC model. Its DM halo was taken to have an NFW profile~\cite{NFW}
\begin{equation}
\rho(r)=\rho_s\left[\frac{r}{r_s}\left(1+\frac{r}{r_s}\right)^2\right]^{-1}
\end{equation}
with scale radius $r_s=$560 kpc and scale density $\rho_s=$0.012 GeV/cm$^3$ \cite{virgohalo}, with a virial mass $M=4.2\times10^{14} M_\odot$, and its distance was taken to be 15.4 Mpc \cite{virgodistance}. While DM halo substructure could in principle increase the flux signal by as much as three orders of magnitude \cite{clustersub}, this would still be subdominant, though it could be brighter than the GC if the Milky-Way halo has a cored DM density profile \cite{fornax}.

In the case that the Universe contains an abundance of halo substructure, the clustering could allow a diffuse extragalactic signal to be visible, through Galactic halo substructure, or extragalactic large scale structure that could be extracted by signal anisotropy analyses \cite{DManisotropy}, or by cross-correlation with large scale structure \cite{DMcorrelation}.

\subsection{Signals from DM annihilations inside the Sun}

Another potential source of neutrino signal is the annihilation of DM particles inside the Sun. DM particles that pass through the Sun lose energy due to DM-nucleon scattering, which is mediated by $Z'$, and hence can become gravitationally trapped~\cite{bib:sun}. Annihilation can sequentially happen as the DM density builds up at the center of the Sun, which produce neutrinos that penetrates the solar medium.
The total rate of annihilation events in the Sun $\Gamma_{\rm ann}$ is given by~\cite{jungman}
\begin{eqnarray} \label{eqAnnihilation}
\Gamma_{\mathrm{ann}} = {C \over 2} {\rm tanh}^2(\sqrt{C A} t) \, .
\end{eqnarray}
Here $C$ is the capture rate of DM particles by the Sun, which is related to $\sigma_{\rm SI}$ in the $B-L$ model, and $A$ is related to the DM annihilation cross section $\sigma_{\rm ann}$ (for details, see~\cite{jungman}).

We note that $\Gamma_{\mathrm{ann}} \approx C/2$ as long as $t > \tau_{\mathrm{eq}} \equiv (\sqrt{C A})^{-1}$. In this case, DM capture by and annihilation in the Sun reach equilibrium, and the total annihilation rate $\Gamma_{\rm ann}$ is set by the capture rate $C$.
For a DM mass above the Icecube energy threshold, the maximal value of $\sigma_{\rm SI}$ allowed by direct detection experiments is $\sigma_{\rm SI} \sim {\cal O}(10^{-9})$ pb~\cite{lux}. Then the nominal value of the annihilation rate $\langle \sigma_{\rm ann} v \rangle \sim 1$ pb generally satisfies the equilibrium condition $\tau_\odot \gtrsim \tau_{\rm eq}$, where $\tau_\odot$ denotes the age of the Sun. This leads to ${\cal O}(10^{19})$ annihilation events per year form the Sun.

The prompt neutrino spectra from the aforementioned $\phi\phi, NN$ annihilation channels, as well as the $t\bar{t}, W^+W^-$ final states from a MSSM-like DM case, are injected at the center of the Sun and propagated to the solar surface, then to the Earth via vacuum oscillation. We have taken full propagation effects into account (for details see Ref.~\cite{Barger:2011em}). The scattering during propagation causes spectral smearing towards to lower energy, as illustrated in Fig.~\ref{fig:nu_spectra} (right panel). The signal spectra from the $t\bar{t},$ and $W^+W^-$ channels can resemble that from the atmospheric neutrinos, making them more difficult to distinguish from the atmospheric background. The harder spectra from $\phi\phi, NN$ channels, however, can be more easily discovered at the IceCube experiment.

Table~\ref{tab:nu_rates} lists the annual (half-year) signal rates at IceCube.  When associated with other indirect searches, the shape difference among the channels can help distinguish the annihilation channels, which we withhold for later discussion in Section~\ref{sect:discussion}. For the calculation of IceCube rates, since the Sun is below the horizon during the half-year polar night, the entire IceCube can be used; we only consider the dominant contained events in the direction of Sun, and assume a 100\% detection efficiency with a 1km$^3$ effective volume. With the maximal LUX allowed $\sigma_n$ and $\left<\sigma v\right>= 1 $pb, the $\phi\phi, NN$ signals are about one order of magnitude below the atmospheric background, and can be readily tested with upcoming IceCube data.

\section{$N_{\rm eff}$ as a Complementary Probe}

The effective number of neutrino $N_{\rm eff}$ plays an important role in identifying the nature of neutrinos (Majorana vs Dirac) in the $U(1)_{B-L}$ model. The information provided by $N_{\rm eff}$ is complementary to that from indirect detection because it is completely independent from the identity of DM in the $B-L$ model.

The important point is that in this model the RH neutrinos $N$ interact with the SM particles via the exchange of $Z^{\prime}$ in this model. As we will see, this implies that if $N$ is a Dirac partner of a LH neutrino $\nu$, it will stay in thermal equilibrium with the primordial plasma down to low temperatures. The presence of one or more Dirac neutrinos can therefore affect $N_{\rm eff}$, which is constrained by CMB experiments~\cite{planck}.

Let us first consider interactions between $N$ and the SM particles in more detail. We focus on charged leptons and LH neutrinos as number density of baryons is suppressed at sub-GeV temperatures, which are of relevance here, and interaction with photons is loop suppressed. $Z^{\prime}$ exchange yields the following effective interactions:
\begin{eqnarray} \label{eff}
{\cal L}_{\rm eff} & \supset & \left({g_{B-L} Q_L \over 2 m_{Z^{\prime}}}\right)^2 2 \left[{\bar \psi}_\nu (1 - \gamma_5) \psi_\nu \right]  \left[{\bar \psi}_l \psi_l \right] \, \nonumber \\
& + & \left({g_{B-L} Q_L \over 2 m_{Z^{\prime}}}\right)^2 \left[{\bar \psi}_\nu (1 - \gamma_5) \psi_\nu \right] \left[{\bar \psi}_\nu (1 + \gamma_5) \psi_\nu \right] \, , \nonumber \\
~ ~
\end{eqnarray}
where $\psi_\nu$ denotes a Dirac neutrino containing $N$ and $\nu$, and $\psi_l$ is a Dirac fermion containing the LH and RH components of a charged lepton.

The two terms on the RH side of Eq.~(\ref{eff}) result in the following cross sections
\begin{eqnarray} \label{cs}
\sigma_{N l \rightarrow N l} \approx \left({g_{B-L} Q_L \over m_{Z^{\prime}}}\right)^4 {s \over 6} \, , \nonumber \\
\sigma_{N \nu \rightarrow N \nu} \approx \left({g_{B-L} Q_L \over m_{Z^{\prime}}}\right)^4 {s \over 24} \, .
\end{eqnarray}
for scattering of a RH neutrino $N$ off a charged lepton $l$ and a LH neutrino $\nu$ respectively. The total rate for interactions that keep $N$ in kinetic equilibrium with the plasma is $\Gamma_{\rm kin} = \langle \sigma_{N l \rightarrow N l} v_{\rm rel} \rangle n_l + \langle \sigma_{N \nu \rightarrow N \nu} v_{\rm rel} \rangle n_\nu$. Here $\langle \rangle$ denotes thermal averaging ($v_{\rm rel}$ being the relative velocity between scattering particles) and $n_l, ~ n_\nu$ are mean number density of charged leptons and LH neutrinos respectively.

Kinetic decoupling of the RH neutrino $N$ occurs at a temperature $T^N_{\rm d}$ when $\Gamma_{\rm kin}$ drops below the Hubble expansion rate of the universe $H = (\pi^2 g_*(T)/30)^{1/2} T^2/M_{\rm P}$, where $T$ is the plasma temperature and $g_*(T)$ is the number of relativistic degrees of freedom at $T$. The contribution of the RH neutrinos to $N_{\rm eff}$ is given by:
\begin{equation} \label{Neff1}
\Delta N_{\rm eff} = \left({g_*(T^\nu_{\rm d}) \over g_*(T^N_{\rm d})}\right)^{4/3} N_{\rm Dirac} ,
\end{equation}
where $T^\nu_{\rm d} \sim {\cal O}({\rm MeV})$ is the decoupling temperature of the LH neutrinos, $g_*(T^\nu_{\rm d}) = 10.75$, and $N_{\rm Dirac}$ denotes the number of Dirac neutrinos.

For $(m_Z^{\prime}/g_{B-L} Q_L) > 6$ TeV, to be compatible with the collider bounds~\cite{Z'}, the decoupling temperature of $N$ is found to be $T^N_{\rm d} > 140$ MeV. At such temperatures new relativistic degrees of freedom become accessible (starting with muons and pions), which implies $g_*(T^N_{\rm d}) > 17.25$, and hence:
\begin{equation} \label{Neff2}
\Delta N_{\rm eff} < 0.53 ~ N_{\rm Dirac} .
\end{equation}

The present observational bound on $\Delta N_{\rm eff}$ from Planck+WMAP9+ACT+SPT+BAO+HST at 2$\sigma$ is $\Delta N_{\rm eff} = 0.48^{+0.48}_{-0.45}$~\cite{planck}, which implies that $\Delta N_{\rm eff} = 0.96$ at 2$\sigma$. If $N_{\rm Dirac} = 3$ (i.e., all three neutrinos are Dirac), $\Delta N_{\rm eff}$ will be in agreement with the current limits provided that $g_*(T^N_{\rm d}) > 25$, see Eq.~(\ref{Neff1}), which requires that $(m_{Z^{\prime}}/g_{B-L}Q_L) \gg 6$ TeV. The situation is more flexible if $N_{\rm Dirac} < 3$. For example, the ``schizophrenic neutrinos'' scenario~\cite{adm}, in which $N_{\rm Dirac} = 1$, is in perfect agreement with the observational constraints as long as  $(m_{Z^{\prime}}/g_{B-L}Q_L) > 6$ TeV.
\vskip 3mm
\noindent
We conclude that $N_{\rm eff}$ can convincingly discriminate between the cases with or without Dirac neutrinos in the $B-L$ model. The conclusion is robust since $\Delta N_{\rm eff}$ is significantly different from zero even if the $Z^{\prime}$ mass is much heavier than the current collider bounds. It is also independent from the identity of DM (${\tilde \chi}^0_1$ or ${\tilde N}$) in this model.

\begin{table}[t]
\scriptsize

\begin{tabular}{c|c|cc|cc}
\hline
Channels&Mid.Lat.&\multicolumn{2}{c|}{Sun}&\multicolumn{2}{c}{GC}\\
\hline
{ $M_{DM}$=200 GeV} & Int.$\gamma$\ &Int.
 $\nu_\mu$/$\gamma$\ \  & evts/yr &Int.
 $\nu_\mu$/$\gamma$\ \  & evts/yr \\
\hline
$\phi\phi$&10.1 &1.0 &9.5 &1.0 &0.28 \\
$NN$ & 42.1 &0.30 &19 &0.14 &0.19 \\
$t\bar{t}$&46.5 &0.079 &3.4 &0.083 &0.088 \\
$W^+W^-$ & 24.5 &0.31 &13 &0.20 &0.16\\
\hline
{$M_{DM}$=300 GeV}& Int.$\gamma$\  &Int.
 $\nu_\mu$/$\gamma$\ \  & evts/yr &Int.
 $\nu_\mu$/$\gamma$\ \  & evts/yr \\
 \hline
$\phi\phi$&10.4 &1.0 &31 &1.0 &0.22 \\
$NN$&43.4  &0.34 &53 &0.13 &0.20 \\
$t\bar{t}$&52.9  &0.060 &10 &0.10 &0.072 \\
$W^+W^-$& 25.7 &0.19 &23 &0.24 &0.12 \\
\hline
{$M_{DM}$=400 GeV} & Int.$\gamma$\ &Int.
 $\nu_\mu$/$\gamma$\ \  & evts/yr &Int.
 $\nu_\mu$/$\gamma$\ \  & evts/yr \\
\hline
$\phi\phi$& 10.7 &1.0 & 44&1.0 &0.18 \\
$NN$& 44.7  &0.22 &69 &0.13 &0.17 \\
$t\bar{t}$& 58.6 &0.051 &13 &0.12 &0.061 \\
$W^+W^-$& 26.5 &0.15 &24 &0.29 &0.10 \\
\hline
{$M_{DM}$=600 GeV} & Int.$\gamma$\ &Int.
 $\nu_\mu$/$\gamma$\ \  & evts/yr &Int.
 $\nu_\mu$/$\gamma$\ \  & evts/yr \\
\hline
$\phi\phi$&10.9 &1.0 &66 &1.0 &0.13 \\
$NN$& 46.8  &0.025 &16 &0.16 &0.13 \\
$t\bar{t}$& 67.7 &0.043 &18 &0.15 &0.048 \\
$W^+W^-$&27.3 &0.14 &28 &0.38 &0.077 \\
\hline
\end{tabular}
\normalsize
\caption{Comparison of integrated spectra within experimental limits and the annual signal event rates. The signal from the Sun and GC assume only the contained events inside the Icecube (1km$^3$) and DeepCore effective volumes, respectively. The propagated prompt $\nu_\mu$ spectra (including oscillation and attenuation effects) are integrated from 70 GeV (IC) and 10 GeV (DC) up to the DM mass, and the prompt $\gamma$ spectra are integrated between 0.2 and 300 GeV. The solar signal rate assumes the maximal DM-nucleon scattering rate allowed by LUX at each DM mass. The `Int. $\nu_\mu/\gamma$' is the ratio of the integrated $\nu_\mu$ and $\gamma$ fluxes, normalized by the value from the $\phi\phi$ channel.
}
\label{tab:nu_rates}
\end{table}

\section{Discussion and Conclusion}
\label{sect:discussion}

We now discuss the results found in the previous sections. In Sec.~\ref{sect:ID}, we found that the astrophysical gamma-ray and neutrino annihilation spectra of $B-L$ dark matter models are distinguishable from typical observable MSSM DM annihilation spectra. We considered the gamma-ray signal from annihilations in a smooth Milky-Way halo, observed at the mid-galactic latitudes $5^\circ<|b|<15^\circ$ and $|l|<80^\circ$. We also considered the flux of neutrinos from annihilations in a $5^\circ$ cone around the Galactic center, and from annihilations in the Sun when DM-nucleon scattering is significant. To gain intuition about the size of these signals, we give the rate of signal events for each of these signals in Columns 2, 4, and 6 respectively in Table~\ref{tab:nu_rates}. The gamma-ray events are integrated from 0.2 to 300 GeV, the GC neutrino events are greater than 70 GeV, and the Sun neutrino events are greater than 10 GeV.

Given the small signal to noise of these signals, it is important to have observable properties of the spectra that are able to distinguish them. One useful measurement to consider is a ratio of detected neutrino to gamma-ray events from two different experiments. This works as a simple way to amplify the differences between the annihilation products. To demonstrate, we integrated the spectra over $10^{-3} m_{\text{DM}}\leq E\leq m_{\text{DM}}$ to determine the incident number of particles, and took the ratio of gamma-rays to neutrinos, shown in Columns~3 and 5 of Table~\ref{tab:nu_rates}, listed with respect to the ratio from the $\phi\phi$ channel. We see that, whether or not prompt neutrinos from $NN$ decay are present, the $B-L\ \phi\phi$ spectra produce a much larger ratio of neutrinos than the observable MSSM spectra. While the specific count and $\nu/\gamma$ ratio magnitudes will depend on the experiments, the relative $\nu/\gamma$ ratio between the different spectra for fixed experiments will remain similar.

This result is robust over a variety of energy ranges, and different neutrino flavor sensitivities, according to each experiment. In Column~2 of Table~\ref{tab:nu_rates}, we integrated the gamma-ray spectrum over $0.2\text{ GeV}\leq E_\gamma\leq300\text{ GeV}$, and the muon-only neutrino flux over $E_{\nu_\mu}\geq10\text{ GeV}$. While other details such as energy-dependent effective areas and viewing strategies will modify the magnitude of each ratio, the relative separation between the different spectra will persist. Thus, the neutrino-philic nature of $B-L$ dark matter is a robust method for distinguishing it from MSSM indirect detection signals.

The one exception to this conclusion is if LH neutrinos are Majorana and the annihilation is dominated by NN. This could happen, for example, if $m_\phi>m_{\text{DM}}$. Then we see that over the energy ranges we considered, the LH neutrinos are produced in similar relative abundance to gamma-rays as in the MSSM models. However, this model is still distinguishable by other features. If the RH N mass is near the DM mass, then the prompt neutrino feature is very prominent. In the other extreme where $m_N\ll m_{\text{DM}}$, then the spectrum is dominated by the decay of the Higgs $h$. In both cases, the kinematic cutoff of the spectrum is set by the maximum energy of the $h$, at a lower energy than the other spectra. This could be established if the DM mass were determined through other means.

If a $\nu$-philic spectrum were observed, suggesting the $B-L$ scenario, then the absence of prompt neutrinos in the $B-L$ dark matter annihilation spectrum would mean that RH neutrino decay is not seen. In the context of our model, this means that either neutrinos are Dirac, or it could be that neutrinos are Majorana and that either $m_N > m_{\text{DM}}$ making annihilation to $NN$ kinematically inaccessible, or $m_N\ll m_{\text{DM}}$ and the prompt neutrino spectrum is washed out over too large an energy range to be observed over the $\phi\phi$ spectrum.

However, Dirac neutrinos must provide additional contributions to $N_{\text{eff}}$, modifying the large scale structure power spectrum, and hence the fluctuations of cosmic microwave background. The scale of the $Z'$ mass determines the decoupling temperature of $N$ and the required $\Delta N_{\text{eff}}$. If this is not observed, then neutrinos must be Majorana.

These results apply equally to $\tilde{N}$ DM, and $B-L\ \tilde{\chi}_1^0$ DM, except that the indirect detection signal from the Sun is not possible for $\tilde{\chi}_1^0$ DM. Likewise, the observation of a direct detection signal is only possible for the $\tilde{N}$ scenario, though it may be out of reach for Majorana neutrinos if the mass separation between the $\tilde{N}$ mass eigenstates is too large.

Table~\ref{tab:signals} summarizes these results. The signals we have discussed are complementary to the observations that would be made in collider experiments, and are an important contribution to determining a globally consistent particle theory.

\begin{table}
\begin{tabular}{l|cc|cc}
\hline\hline
\centering{Nature of Neutrino} & \multicolumn{2}{|c|}{Majorana $\nu$} & \multicolumn{2}{c}{Dirac $\nu$} \\
\hline
$B-L$ Sector Candidate of Dark Matter & \quad$\tilde{N}$\quad\mbox{} & \quad$\tilde{\chi}_1^0$\quad\mbox{} & \quad $\tilde{N}$\quad\mbox{} & \quad$\tilde{\chi}_1^0$\quad\mbox{} \\
\hline
$\nu$-philic Annihilation Spectrum?  & Yes & Yes  & Yes & Yes \\
Prompt $\nu$ Annihilation Spectrum?& Maybe & Maybe  & No & No \\
$\nu$ Signal from Sun/Direct Detection?  & Maybe & No  & Maybe & No \\
Non-SM Contributions to $N_\text{eff}$?  & No & No  & Yes & Yes \\
\hline\hline
\end{tabular}
\caption{Summary of the signals that distinguish $B-L$ dark matter models and the nature of neutrinos.}
\label{tab:signals}
\end{table}

We finish this Section by pointing out that, although these results were found in the context of a gauged $U(1)_{B-L}$-extended MSSM, many of the features discussed here are generic for any model. If DM annihilation occurs through exchange of neutrinos or their superpartners, then the presence or absence of mass-flipped contributions will distinguish between Majorana or Dirac neutrinos. If neutrinos are Dirac and RH neutrinos remain coupled to the primordial plasma until late times, then their contributions to $N_\text{eff}$ must be observed.

In conclusion, in this paper we investigated the annihilation of DM into photons and neutrinos in the context of a well-motivated model that explains small neutrino mass. The model possesses a $U(1)_{B-L}$ gauge symmetry. The DM candidate in such a scenario can be either the lightest neutralino in the $B-L$ sector or the lightest RH sneutrino. The final states arising from the DM annihilation can contain lot of neutrinos that will help distinguish this model from the MSSM at IceCube. The neutrinos in the final state arise from $S$-wave dominated annihilation of DM.

We also showed that it may be possible to ascertain whether the light neutrinos have a Dirac or Majorana nature. When the light neutrinos are Majorana type, then the spectrum may show a prompt neutrino box-feature and in addition, the spectrum may exhibit a lower energy kinematic cut-off. The gamma ray signal from the galactic center may show an early cut-off in such a scenario. When the light neutrinos are Dirac type, we may find their contributions in the measurement of $N_{\text{eff}}$ from the cosmic microwave background data. The signal at IceCube depends on the direct detection cross-section (capture rate) and the annihilation cross-section for DM equilibration in the Sun. Any near-future observation of the neutrinos from the $B-L$ model at IceCube will promise a discovery of this model at the direct detection experiments very soon. The larger annihilation cross-section, as required for equilibration, will make the future detection of the signal from the galactic center hopeful.



\section{Acknowledgements}

S.S.C. thanks Ranjan Laha for helpful discussions. The work of B.D. is supported by DOE Grant DE-FG02-13ER42020. S.S.C. is partially supported by NSF Grant PHYS-1101216 to John F. Beacom. Y.G. thanks the Mitchell Institute for Fundamental Physics and Astronomy for support. R.A. and B.D. acknowledge partial support by the National Science Foundation (NSF) under Grant No. PHYS-1066293 and the hospitality of the Aspen Center for Physics where this work began.

\end{document}